\documentclass{nature}
\usepackage{eqnarray}
\usepackage[english]{babel} 
\usepackage[]{layout}
\usepackage{amsmath,amsfonts,amssymb}
\usepackage{graphicx}
\usepackage{float}
\usepackage{color}
\usepackage{braket}
\usepackage{array,multirow,makecell}
\makeatletter
\let\saved@includegraphics\includegraphics
\AtBeginDocument{\let\includegraphics\saved@includegraphics}
\renewenvironment*{figure}{\@float{figure}}{\end@float}
\makeatother

\title{Nontrivial band geometry in an optically active system}

\author{Jiahuan Ren$^{1,2}$, Qing Liao$^{1\ast}$, Feng Li$^{3,4\ast}$, Yiming Li $^{3}$, Olivier Bleu$^{5}$, \\ Guillaume Malpuech$^{5}$, Jiannian Yao$^{2}$, Hongbing Fu$^{1,2\ast}$, Dmitry Solnyshkov$^{5\ast}$}

\begin{document}
\maketitle

\begin{affiliations}
\item Beijing Key Laboratory for Optical Materials and Photonic Devices, Department of Chemistry, Capital Normal University, Beijing 100048, People's Republic of China
\item Tianjin Key Laboratory of Molecular Optoelectronic Sciences, Department of Chemistry, School of Sciences, Tianjin University, Collaborative Innovation Center of Chemical Science and Engineering, Tianjin 300072, People's Republic of China
\item Key Laboratory for Physical Electronics and Devices of the Ministry of Education \& Shaanxi Key Lab of Information Photonic Technique, School of Electronic and Information Engineering, Xi'an Jiaotong University, Xi'an 710049, People's Republic of China
\item Department of Physics and Astronomy, University of Sheffield, Sheffield, UK
\item Institut Pascal, PHOTON-N2, Universit\'e Clermont Auvergne, CNRS, SIGMA Clermont, F-63000 Clermont-Ferrand, France.
\item Institut Universitaire de France (IUF), 75231 Paris, France
\end{affiliations}

\begin{abstract}
Optical activity (OA), also called circular birefringence, is known for two hundred years, but its applications for topological photonics remain unexplored. Unlike the Faraday effect, OA provokes rotation of the linear polarization of light without magnetic effects, thus preserving the time-reversal symmetry.  Here, we report a direct measurement of the Berry curvature and quantum metric of the photonic modes of a planar cavity containing an optically active organic microcrystal (perylene). Photonic spin-orbit-coupling induced by the cavity results in the action of a non-Abelian gauge field on photons. The addition of high OA makes emerge geometrically non-trivial bands containing two gapped Dirac cones with opposite topological charges. This experiment performed at room temperature  and at visible wavelength establishes the potential of optically active organic materials for implementing non-magnetic and low-cost topological photonic devices.
\end{abstract}

\maketitle

The exploration of photonic systems involving the concepts of topology  and non-reciprocity has become a mainline of scientific activity in recent years, driven by both fundamental and applied motivation. Indeed, the implementation of microscopic optical isolators is absolutely crucial for the development of integrated photonics \cite{Stadler2014} and robust quantum optical circuits \cite{Barik2018}. Photonic topological insulator analogs \cite{Haldane2008,lu2014topological,Ozawa2019} and topological lasers \cite{StJean2017,Bahari2017,Bandres2018,klembt2018exciton,Mittal2018} with topological edge modes represent a solution to this stringent request. There are two great families of 2D topological insulators \cite{Hasan2010}. One is based on the Quantum Hall effect (QHE), either normal or anomalous (QAHE), in systems with broken time-reversal symmetry (TRS). The energy bands possess non-zero integrated Berry curvature (Chern numbers), which leads to non-reciprocal transport on the edges. The other family is based on the Quantum Spin Hall effect (QSHE). The total band Chern number is zero, but the band geometry remains non-trivial. In particular, it is possible to separate two spin or pseudo-spin domains, each being characterized by a non-zero Berry curvature, opposite between the two. The integration over the (pseudo)-spin subbands gives non-zero (pseudo)-spin Chern numbers. If the corresponding (pseudo)-spins are protected by a symmetry, such as the TRS, which protects the electron's spin, non-reciprocal (pseudo)-spin transport on the edge or interface states can take place. This type of effect has been demonstrated in photonics with  various types of pseudo-spin realizations, each being approximately protected by a specific symmetry \cite{Hafezi2013,Ma2015,Cheng2016,Khanikaev2017,Gao2018}.

Photonic QAHE requires the combination of photonic spin-orbit coupling (SOC) \cite{Kavokin2005}, which is an intrinsic property of 2D confined photonic media like waveguides and planar cavities, with TRS breaking by the Faraday effect \cite{Faraday1845} induced by an applied magnetic field. Because of these two contributions, the photonic modes of a 2D continuous medium exhibit a non-zero Berry curvature \cite{Bleu2018effective,Gianfrate2019}. 
Once inserted in an appropriate 2D lattice, these modes demonstrate topological gaps and non-reciprocal transport on the lattice edge \cite{Haldane2008,Wang2009,lu2014topological,Nalitov2015b,Bahari2017,klembt2018exciton}. 
However, the Faraday effect is usually small at optical wavelengths. It requires large magnetic fields, hindering practical applications. The so-called optical activity (OA) is another type of optical response discovered at the beginning of the XIXth century \cite{Arago1811,Pasteur1848}, leading (similar to the Faraday effect) to the rotation of the linear polarization of light during its propagation. Unlike the Faraday effect, OA is an intrinsic property of a crystal or of a molecule linked with their chirality. It does not require magnetic field and preserves the TRS, as sketched in Fig.~1(A). In the Faraday effect, the angle of rotation of the polarization $\alpha_0$ continues to increase for inverted propagation direction. With OA, the polarization starts to rotate backwards to its original position, thus preserving the TRS. Perylene \cite{Donaldson1953} (C$_{20}$H$_{12}$) under its crystallized form is a specific organic microcrystal showing strong OA due to its crystal structure. Perylenes are nowadays widely used in optics due to its intense light absorption and luminescence in the visible range, high stability and quantum yield \cite{Huang2011}. Their optical properties are promising for solar cells  \cite{Meng2018}, energy harvesting, temperature control \cite{Russ2016}, lasers \cite{Rulliere1972}, and other nanophotonic applications \cite{Weil2010}. Perylenes are also used in the field of 2D materials to hold together van der Waals heterostructures \cite{Novoselov2016} or by themselves \cite{Jariwala2017}.

\begin{figure}[tbp]
\includegraphics[width=1\linewidth]{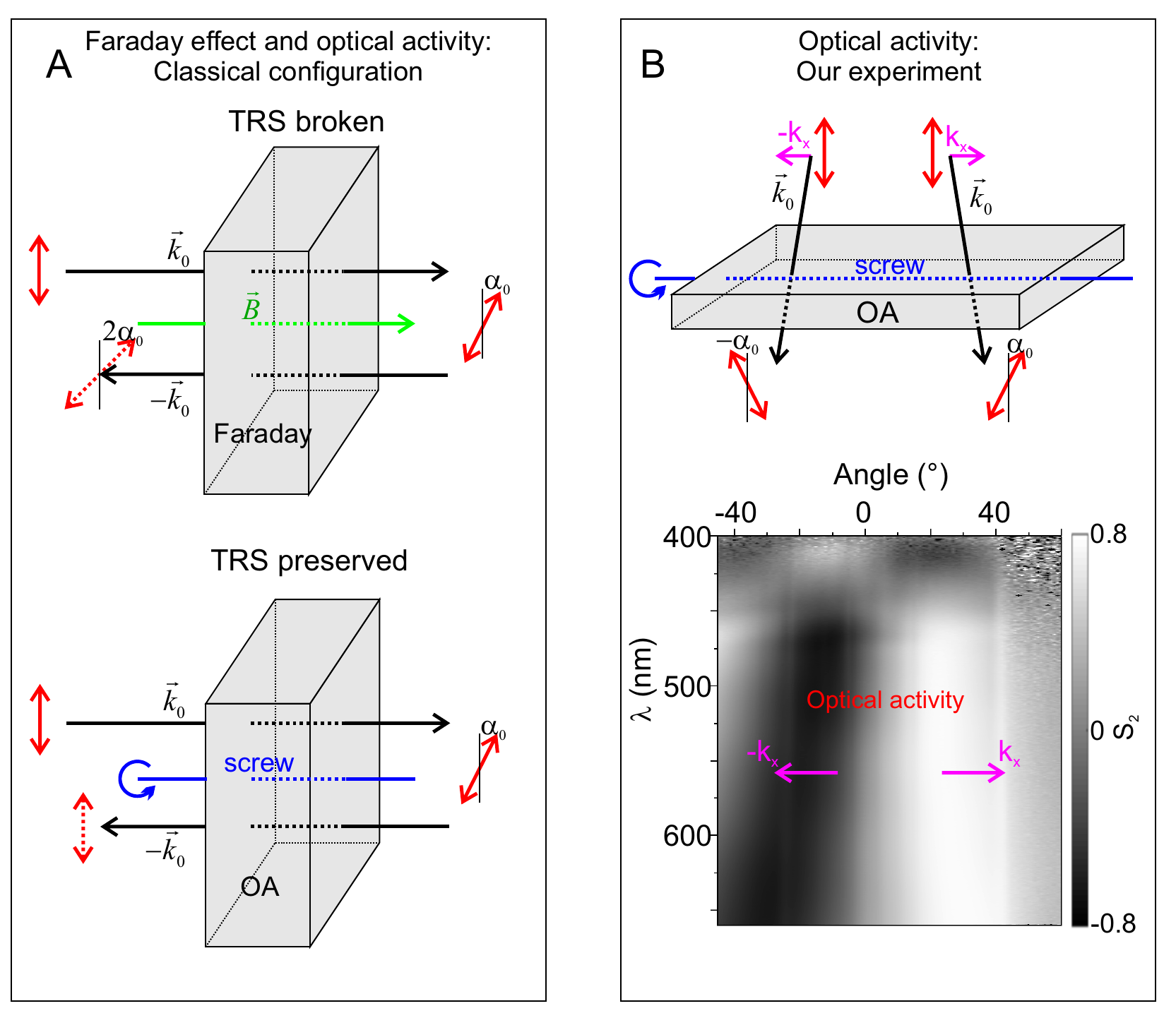}
\caption{\textbf{Faraday effect and optical activity.} A) Scheme showing the rotation $\alpha_0$ of the linear polarization (red double arrow) of light while making a back and forth trip, in the Faraday effect (top, TRS broken by magnetic field $\vec{B}$ -- green) and in a chiral OA material (bottom, TRS not broken, screw axis in blue). Black arrow indicate the propagation direction (the light wave vector $\vec{k}_0$). B) Optical activity in perylene: (top) scheme of the experiment demonstrating opposite rotation of the linear polarization for opposite in-plane wave vector $k_x$. The screw axis is shown in blue, and the propagation wave vector $\vec{k}_0$ is shown as a black arrow. (bottom): Measured diagonal polarization degree (Stokes vector $S_2$ determined by the linear polarization rotation due to the OA in perylene.}
\end{figure}

In this work, we establish the potential of OA media for topological photonics. We study a basic photonic element -- a planar cavity, filled with an OA perylene crystal. We provide a direct measurement of the complete band geometry, namely the Berry curvature and the quantum metric of the photonic bands. We show that their non-trivial distribution can be interpreted as the result of the action of a non-Abelian gauge field, which itself results from the interplay between the photonic spin-orbit coupling induced by the optical cavity and the optical activity.

\section{Results}

We start by illustrating the OA of perylene in Fig.~1(B). The crystallized organic microcrystal is excited by linearly polarized light whose polarization is aligned with the crystal axes and the diagonal polarization degree (proportional to the rotation angle $\alpha_0$) is measured in transmission versus energy and wave vector $k_x$ (see Materials and Methods for details). 
This experiment demonstrates the key feature of OA, which is that the sense of rotation of the linear polarization is given by the sign of $k_x$. While the space group symmetry of perylene $2_1/m$ does not allow optical activity because of the glide plane, the delocalized nature of the excitons in this crystal \cite{Rangel2018} removes this constraint (see Methods). The origin of the effect is confirmed by the change of sign of the rotation at the excitonic resonance. However, a precise quantitative measurement of the optical activity of a pure crystal remains difficult, as usual, because of the dominating linear birefringence. 

We aim to compensate the linear birefringence and to study the optical activity by embedding the perylene crystal within a metallic microcavity (Fabry-Perot resonator)  as sketched in Fig.~2(A). The quantization of the z-component of the wave vector leads to the formation of 2D bands parametrized by the in-plane 2D wave vector $\mathbf{k}=\left(k_x,k_y\right)^T$.
The modes we consider are above the light cone. They have small in-plane wavevector, parabolic dispersion, and are radiatively coupled to the outside of the cavity \cite{Microcavities}.
The experimental setup (see Supplementary Materials for details), allowing to make a full optical state tomography out of which the band geometry (Berry curvature and quantum metric) can be reconstructed \cite{Bleu2018effective}, is shown in Fig.~2(B). All experiments are performed at room temperature. The reflection of the sample is measured versus energy and in plane wave vector for the 6 different light polarizations (left and right circular, horizontal-vertical, and diagonal-antidiagonal), which allows a full determination of the three Stokes vector components (light polarization pseudo-spin)  of the modes.
The axes of the polarizer are aligned with the crystal axes and with the axes of the reciprocal space defined by the CCD camera. The orientation of the Stokes vector $\mathbf{S}(\mathbf{k})$ on the Poincar\'e sphere is given by the polar angle $\theta(\mathbf{k})$ and the azimuthal angle $\phi(\mathbf{k})$. The measurement of these quantities allows to extract the Berry curvature\cite{berry1984quantal} $B_z$  and the quantum metric \cite{provost1980riemannian}  $g_{ij}$ as \cite{Bleu2018}: 
\begin{eqnarray}
g_{ij} =\frac{1}{4}(\partial_{k_i}\theta\partial_{k_j}\theta+\sin^2\theta \partial_{k_i}\phi\partial_{k_j}\phi) \nonumber\\
B_z =\frac{1}{2}\sin\theta(\partial_{k_x}\theta\partial_{k_y}\phi-\partial_{k_y}\theta\partial_{k_x}\phi)
\label{xtract}
\end{eqnarray}

\begin{figure}[tbp]
\includegraphics[width=1\linewidth]{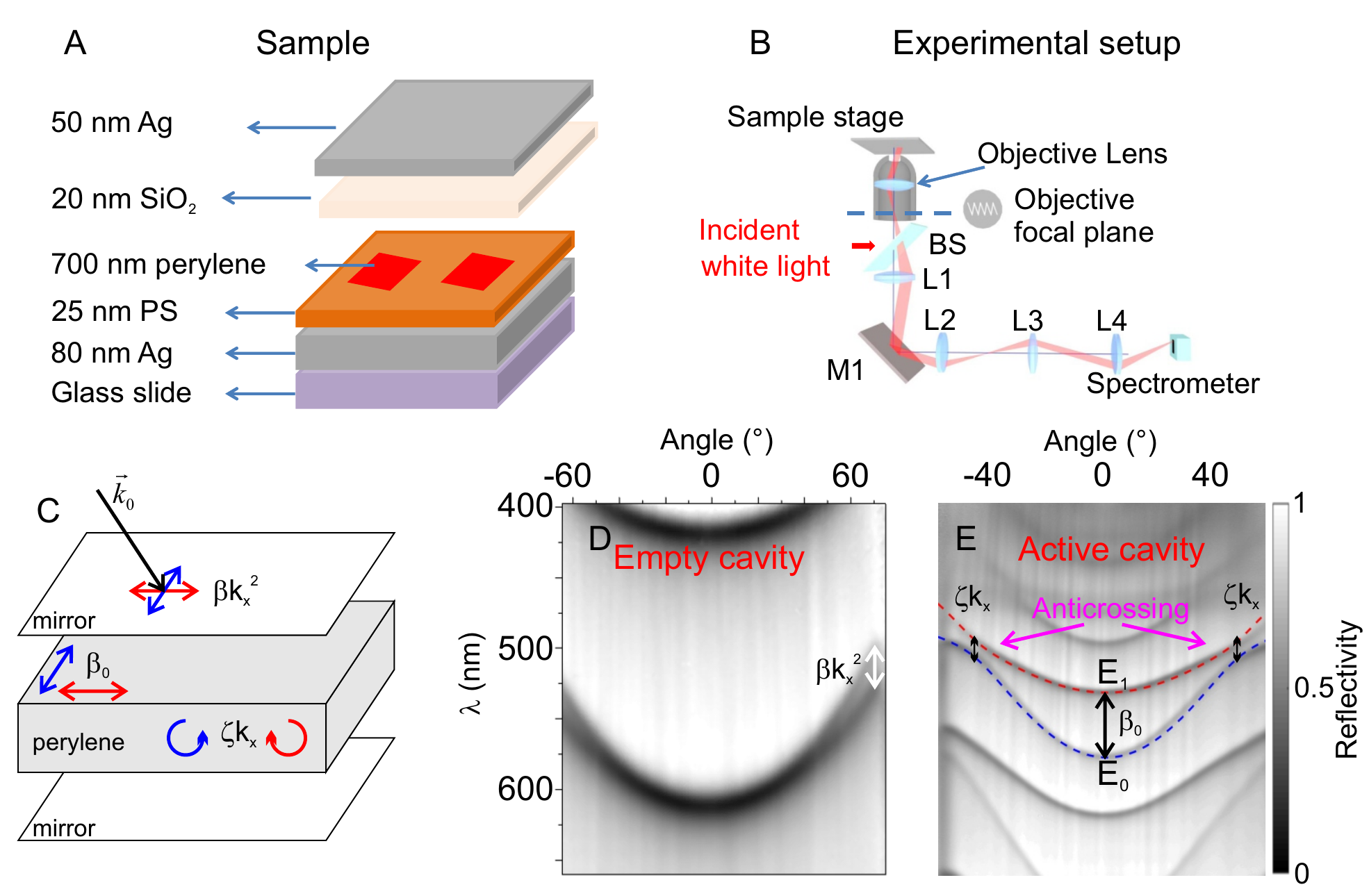}
\caption{\textbf{Combining the spin-orbit couplings.} A) The sample consists of a perylene crystal embedded in a microcavity. B) Experimental setup allowing to obtain polarization-resolved complete state tomography. BS: beam splitter; L1-L4: lenses; M1: mirror. The red beam traces the optical path of the reflected light from the sample at a given angle. C) Embedding the perylene crystal in a planar cavity combines TE-TM splitting $\beta k^2$, OA (circular birefringence) $\zeta k_x$, and linear birefringence $\beta_0$. D) Reflection of an empty cavity versus wavelength and incidence angle which evidences the k-dependent TE-TM splitting $\beta k^2$. E) Reflection of a cavity filled with perylene, which is inducing both linear $\beta_0$ and circular birefringence $\zeta k_x$ (black arrows) on top the cavity-induced TE-TM splitting. Dashed lines show a fit of the two coupled bands with the Hamiltonian (4). The axes of the polarizer and the excitation direction $k_x$ are aligned with the fast axis of the crystal.}
\end{figure}

Embedding the perylene crystal in a cavity results in a combination of optical effects. Indeed, in an ideal case the above-mentioned quantized modes of the cavity are doubly polarization degenerate at zero in-plane wavevector. This polarization degeneracy is lifted by different contributions sketched in Fig.~2(C). The TE-TM splitting of the bare cavity \cite{Panzarini99,Kavokin2005} characterized by $\beta$ is zero at $k=0$ and then grows quadratically with $k$ \cite{Panzarini99,Kavokin2005}:
\begin{equation}
    E_{TE,TM}=\frac{\hbar k^2}{2m_{TE,TM}}=\frac{\hbar^2 k^2}{2m}\pm\beta k^2
\end{equation}
with $m_{TM}$ and $m_{TE}$ corresponding to the longitudinal and transverse effective masses. 
The linear birefringence \cite{Krizhanovskii2003} of the perylene crystal described by $\beta_0$ splits the linearly polarized modes H and V at $k=0$: $E_{H,V}(k=0)=\pm\beta_0$. These terms cancel each other at two points along the $k_x$ axis. Together, they determine the linear polarization of the modes ($S_1$, $S_2$ Stokes components). On the other hand, the OA (circular birefringence, controlled by $\zeta$)
in a cavity leads to a linear in $k_x$ splitting between the circular-polarized modes (see Methods):
\begin{equation}
    E_{\pm}=E_0\pm\zeta k_x
\end{equation}
It therefore determines the circular polarization degree of the modes ($S_3$). 
All these effects can be combined in a single effective $2\times 2$ Hamiltonian describing for one confined mode of the cavity the two polarization eigenstates. We write it on the circular polarization basis:

\begin{equation}
{H_{\bf{k}}} = \left( {\begin{array}{*{20}{c}}
{\frac{{{\hbar ^2}{k^2}}}{{2{m^*}}} + {\zeta}{k_x}}&{\beta_0 + \beta {k^2}{e^{2i\varphi }}}\\
{\beta_0 + \beta {k^2}{e^{ - 2i\varphi }}}&{\frac{{{\hbar ^2}{k^2}}}{{2{m^*}}} - {\zeta}{k_x}}
\end{array}} \right)   
\end{equation}
where $m^*=m_{TM}m_{TE}/(m_{TM}+m_{TE})$ and $\varphi$ the polar angle. As any $2\times 2$ Hermitian Hamiltonian, it is a linear combination of Pauli matrices that can be physically interpreted as an effective magnetic field acting on the Stokes vector. The effective field reads: 
\begin{equation}
{\bf{\Omega }}({\bf{k}}) = \left( {\begin{array}{*{20}{c}}
{\beta_0  + \beta {k^2}\cos 2\varphi }\\
{ - \beta {k^2}\sin 2\varphi }\\
{{\zeta}{k_x}}
\end{array}} \right)
\end{equation}
As defined above, $\beta_0$, $\beta$, and $\zeta$ determine the strength of the effective fields corresponding respectively to linear birefringence, the k-dependent TE-TM splitting and circular birefringence which can be viewed as an effective Zeeman splitting. The k-dependent effective fields can both be interpreted as photonic SOCs. The Stokes vector of the eigen-modes is either aligned or anti-aligned with the effective field $\mathbf{\Omega}$. This picture allows to find these modes easily and to predict their Berry curvature and quantum metric.

Figure~2(D) shows the total reflection coefficient of an empty cavity with bare TE and TM modes with different effective masses because of red$\beta$, but degenerate at $k=0$.  The cavity filled with an OA material [Fig.~2(E)] shows a radically different mode dispersion. The first visible consequence is that the cavity is optically thicker due to the perylene crystal refractive index ($n \approx 2$). More importantly, the modes get split and linearly polarized at $k=0$ because of the linear birefringence of perylene $\beta_0$. These modes anticross at a finite wave vector (instead of simply crossing) because of the circular birefringence $\zeta$ linked with the optical activity coefficient: $\alpha\approx\zeta k_x n^2/2\hbar c$ (also due to perylene, see Supplementary materials for more details). The key difference with respect to a cavity with TRS broken by the  Faraday effect \cite{Gianfrate2019} is that here, the effective Zeeman field changes sign with $k_x$. This Hamiltonian shows two gapped tilted Dirac cones at the two reciprocal space points where the in-plane components of the field cancel. The sign of the mass term, opposite for the two cones, is given by the sign of the effective Zeeman field. 

These anticrossing points are extremely favorable for the measurement of the OA. Indeed, for arbitrary direction in a general crystal, the optical rotation is dominated by linear birefringence. Precise measurements of the OA in such structures can only be carried along optical axes, while for arbitrary directions they are particularly difficult, as shown by the archetypal example of tartaric acid \cite{Mucha1997}. The possibility to cancel the linear birefringence of a crystal $\beta_0$ by the TE-TM splitting of the cavity $\beta k^2$ for a certain $k$ allows us to obtain a precise measurement of the OA ($\zeta$) for this particular direction. If needed, the cavity properties could be modified in order to change $\beta$ and thus obtain the anticrossing at a different wave vector. 

The validity of the effective Hamilonian is confirmed by the measured 2D wave vector maps of the Stokes vector components of the lower branch, shown in Fig.~3(A-C). As expected, the linear birefringence is compensated by the k-dependent TE-TM field at the anticrossing points \cite{Tercas2014}. The two components $S_1$ and $S_2$ of the Stokes vector cancel and change sign around these points, forming 2D monopoles. The third Stokes vector $S_3$ component is maximal at these two points but is changing sign at $k_x=0$ because of the TRS. The cross-sections of the dispersion close to the anticrossing points together with the pseudospin orientation for the two branches are shown in Supplementary Figure~1.

\begin{figure}[tbp]
\includegraphics[width=1\linewidth]{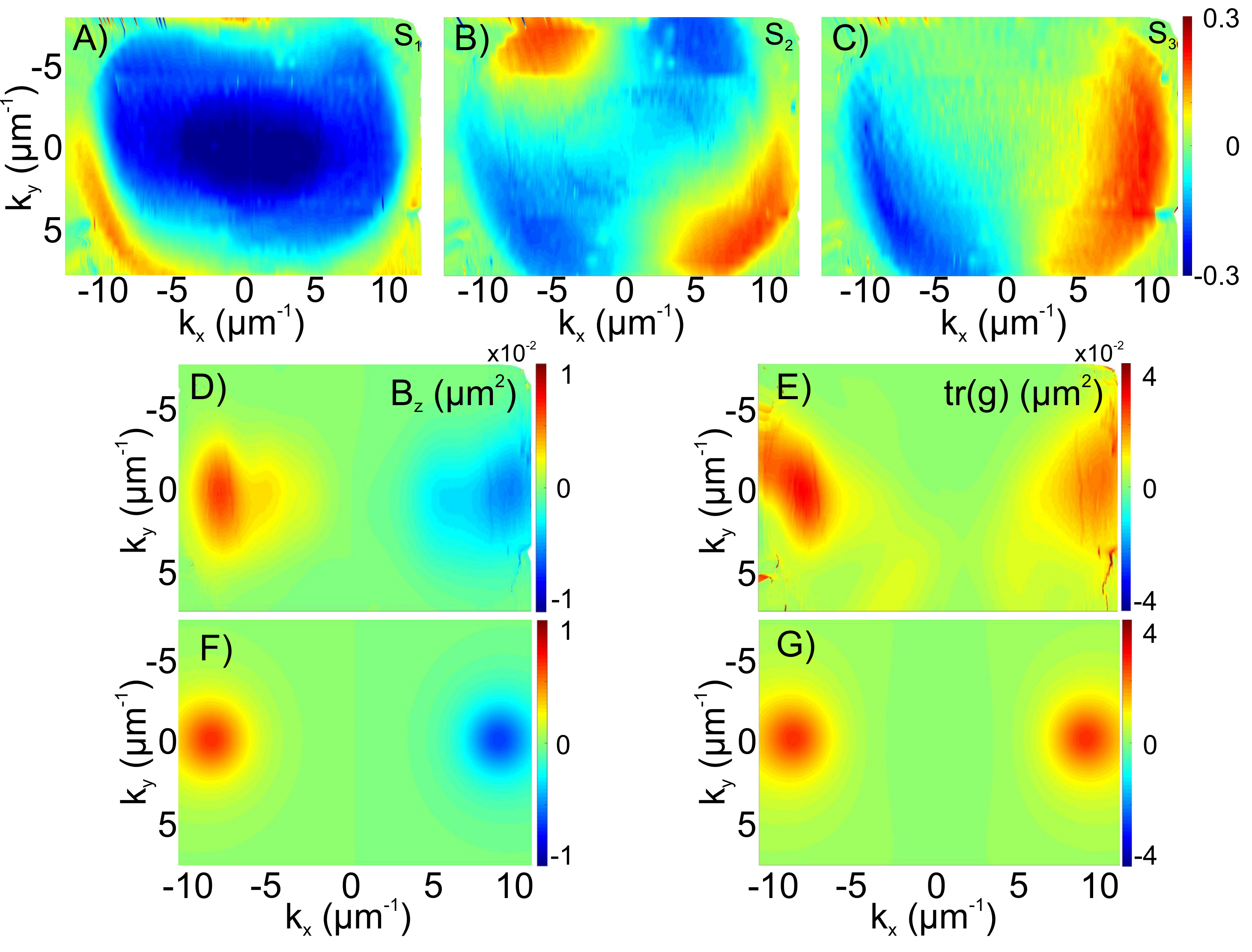}
\caption{\textbf{Stokes parameters, Berry curvature and quantum metric.} Measured Stokes parameters of the mode $E_0$ from Fig.~2E: A) $S_1$, B) $S_2$, C) $S_3$; Quantum geometry extracted from the measured Stokes vector: D)  Berry curvature $B_z$, E) trace of the quantum metric $g_{xx}+g_{yy}$; Calculated quantum geometry based on the mode dispersion: F)  Berry curvature $B_z$, G) trace of the quantum metric $g_{xx}+g_{yy}$.}
\end{figure}

The measurements of the Stokes vector allow to extract the Berry curvature and the quantum metric of the modes, as shown in Fig.~3(D,E). As expected from the TRS (encoded in the Zeeman SOC and the $S_3$ texture), the Berry curvature shows two maxima of opposite signs, which means that the integrated Berry curvature over the band is zero. However, by separating the reciprocal space in two regions, like in the quantum valley Hall effect \cite{Gao2018}, it is possible to associate non-zero pseudo-spin Chern numbers to each of these two regions. These regions can viewed as being analogs of valleys, which emerge in this optically active system without the need of using a lattice. The trace of the quantum metric [Fig.~3(E)] is maximal in the regions of the anticrossing, where the Stokes vector rotation is the fastest. The experimentally extracted geometry (D,E) corresponds very well to the theoretical predictions (F,G) based on the $2\times 2$ effective Hamiltonian (1) with parameters $\beta_0=0.18$ eV, $\beta=9\times 10^{-4}$~eV$\mu$m$^2$, $\zeta=2.5\times10^{-3}$ eV$\mu$m, extracted from the experimental dispersion (Fig.~2(E)). The OA coefficient obtained from the splitting at the anticrossing point demonstrates a remarkably high value of $\alpha=1.4\times 10^4$~degrees/mm. This is a crucial ingredient which has allowed room temperature measurements. Indeed, for a typical OA material such as the tartaric acid, the mode splitting at the anticrossing point would be of the order of $10^{-4}$~meV, much smaller than the broadening $k_BT_{RT}\approx 26$~meV. The room temperature operation at optical wavelengths is highly favourable in the prospect of using such non-trivial band geometry for implementing practical topological photonic devices.

\section{Discussion}

While our study belongs to the field of classical optics, since we are dealing with classical photonic beams, it can nevertheless have important implications for quantum mechanics. Indeed, it is well known that the transverse behavior of a light beam in the paraxial approximation is well described by the Schr\"odinger equation, with the propagation axis playing the role of time. The same applies to planar cavities, but with the meaning of the temporal axis restored. Our system represents therefore a model of a quantum system in many senses. The Stokes vector of light is an equivalent of the spin of an electron, and the non-zero Berry curvature of both the Poincar\'e sphere and the Bloch sphere is the most direct consequence of this analogy. The degrees of freedom provided by the direct and the reciprocal space are also equivalent to quantum mechanics, which allows to use both the languages of Maxwell and Schrodinger equations for topological photonics.

The original antisymmetric distribution of the Berry curvature that we have observed for an OA system crucially affects the numerous phenomena driven by the band geometry, the emblematic one being the anomalous Hall effect family which includes valley Hall effects at the heart of valleytronics \cite{mak2014valley}. With broken TRS (same-sign Berry curvature), the anomalous Hall drift does not change sign upon time reversal (like the Faraday rotation angle $\alpha_0$). With conserved TRS (opposite Berry curvature for $\pm k_x$), the anomalous Hall drift is reversed and the system returns to its original position upon time reversal. These valleys can be selectively excited by simply controlling the beam incidence angle. The precision of such control is very high, since each valley spans about 10~degrees, while a typical beam spans about 4~arc minutes.

Another important quantity which our measure allows to access is the quantum metric \cite{provost1980riemannian}. Quantum metric has been recently found to be associated with many phenomena and it became a hot research topic \cite{Kolodrubetz2017}. In optics, it allows to quantify the non-adiabaticity of realistic transport experiments \cite{Bleu2018effective}, which is certainly crucial to operate devices based on geometrically non-trivial bands such as valleytronic or opto-valleytronic systems. The large value of OA provides a strong protection against the non-adiabaticity, allowing to use very high spatial gradients: the maximal anomalous Hall drift of 0.6~$\mu$m for our parameters can be achieved at a propagation distance of only 45~$\mu$m. 

One more interesting outcome of our work is the possible implementation of an artificial magnetic field acting on photons. Indeed, the effect of the OA in the Hamiltonian (1) can be represented as an action of a vector potential, opposite for the two spin components ($\pm$)
\begin{equation}
    \hat{H}^{\pm}=\frac{1}{2m}\left(\hat{\mathbf{p}}-e\mathbf{A}^{\pm}\right)^2
\end{equation}
where the vector potential is given by $A_x^{\pm}=\pm m\zeta/e$. Making the in-plane OA position-dependent, for example $\zeta(y)=\zeta_0 y$, leads to the emergence of a magnetic field $\mathbf{B}^{\pm}=\nabla\times\mathbf{A}^{\pm}$ of opposite sign for the two circular polarizations. This field gives rise to the formation of Landau levels or Harper-Hofstadter-like energy spectra in periodic structures. This pseudospin-dependent synthetic magnetic field is similar in spirit to the one realized in \cite{Hafezi2013}, but with the use of the  intrinsic polarization pseudospin, which does not require the presence of an artificial lattice. In order to induce such spatial dependence, one can introduce a slight variation of the background refractive index in the cavity, which then affects the $\zeta$ contribution in the Hamiltonian~(1).

Our results show that the organic microcrystals, such as perylene, which are promising for the implementation of classical photonic elements because of their remarkable basic optical properties, are also appealing to implement topological photonic devices, because such devices do not necessarily require broken TRS \cite{BRedondo2018}. The polarization tomography measurements allowed us to fully characterize the quantum geometry of photonic bands. These measurements have revealed that one specific property of this organic microcrystal, optical activity, renders the TRS photonic bands geometrically non-trivial, exhibiting gapped Dirac cones with non-zero Berry curvature. The precise measurements of the quantum geometry of these bands favor the development of quantitative optovalleytronics.

\section{Methods}
\textit{Material structure and fabrication.}
The perylene ($99\%+$) and TBAB (Tetrabutylammonium bromide) used in the experiment were purchased from Acros and Innochem respectively without further purification. The 2D square sheet of crystalline perylene was prepared by space-confined strategy \cite{Wang2018}: $50\mu$L of $0.5$mg/ml perylene/chlorobenzene was first added onto $1$mg/ml TBAB/water solution. After the complete evaporation of chlorobenzene, 2D square sheets of crystalline perylene were formed on the solution surface, exhibiting a thickness of 200 nm-1000 nm. The molecule arrangement of the perylene film is illustrated in Supplementary Figure~2.

\textit{Cavity structure.}
For the empty microcavities characterized by Fig. ~2(d), $80nm$ silver film was first evaporated on a glass substrate, followed by a spin-coating of 300 nm polystyrene (PS) film and a final vacuum evaporation of 50 nm silver film. The reflectance of the $80$nm and $50$nm silver films were $99.4\%$ and $87\%$ respectively, enabling easier light extraction from the top mirror of the cavity. For the active microcavity embedding perylene, $25$nm PS film was first spin-coated on $80$nm silver film, then the prepared 2D perylene sheets (thickness $\sim 750$nm for the studied cavity) were transferred onto the PS film by bringing them in contact at the surface of the TBAB/water solution. A final evaporation of $20$nm SiO$_{2}$ and $50$nm silver film was made to form the microcavity sketched in Fig.~2A.

\textit{Spectroscopy.}
The angle-resolved spectroscopy was performed at room temperature by the Fourier imaging using a $100\times$ objective lens of a NA $0.95$, corresponding to a range of collection angle of $\pm70^\circ$. As sketched in Fig.~2B, an incident white light from a Halogen lamp was focused on the area of the microcavity containing perylene, and the k-space or angular distribution of the reflected light  was located at the back focal plane of the objective lens. Lenses L1-L4 formed a confocal imaging system together with the objective lens, by which the k-space light distribution was first imaged at the right focal plane of L2 through the lens group of L1 and L2, and then further imaged, through the lens group of L3 and L4, at the right focal plane of L4 on the entrance slit of a spectrometer equipped with a liquid-nitrogen-cooled CCD. The use of four lenses here provided flexibility for adjusting the magnification of the final image and efficient light collection. Tomography by scanning the image (laterally shifting L4) across the slit enabled obtaining spectrally resolved 2D k-space images. In order to investigate the polarization properties, we placed a linear polarizer, a half-wave plate and a quarter-wave plate in front of spectrometer to obtain the polarization state of each pixel of the k-space images in the horizontal-vertical ($0^\circ$ and $90^\circ$), diagonal ($\pm45^\circ$) and circular ($\sigma^{+}$ and $\sigma^{-}$ ) basis \cite{Dufferwiel2015, Manni2013}. 

For the spectroscopy on bare perylene shown in Fig.~1B, transmission measurement was used instead due to the low reflectivity of the material. The incident white light was initiated to be linearly polarized before focusing onto the perylene with a large range of incident angles, and the transmitted spectra were imaged in k-space and resolved in polarization. The dependence of polarization on k, featured by the diagonal Stokes parameter $S_{2}$, reveals the nature of OA in the material.

\textit{Extraction of the Stokes parameters and the quantum geometry.}
The experimental tomography images represent a set of intensities in 6 polarization components measured in reflection as a function of in-plane wave vector $(k_x,k_y)$ and wavelength $\lambda$. To obtain the maps of the Stokes vector components for a given mode, shown in Fig.~3A-C, we proceed as follows. For each in-plane wave vector, we first determine the wavelength $\lambda_0$ corresponding to the particular mode, by fitting the total reflection spectrum with Gaussian-broadened resonances over an approximately linear background. We then fit the individual intensity components to determine the relative weight of resonance (taking into account the magnitude and the width of the peak) in each of the 6 polarizations, which allows finally to determine the 3 components of the Stokes vector.

The quantum geometric tensor components $g_{ij}$ and $B_z$ are extracted from the Stokes vector according to Eq.~(3) of the main text. Lowpass Fourier-transform smoothing is applied to the maps of the angles $\theta$ and $\phi$ before calculating the partial derivatives numerically.

\emph{Optical activity.} Different formalisms have been developed for the description of the optical properties of crystals. We start with the equation of the optical indicatrix or the index ellipsoid \cite{Landau8,Newnham2005}, obtained from the relations between $\mathbf{D}$ and $\mathbf{E}$ in the medium, which for a crystal with indices $n_1$ and $n_2$ in the absence of the OA reads
\begin{equation}
    \left(n^2-n_1^2\right)\left(n^2-n_2^2\right)=0
\end{equation}
where $n$ is the refractive index for a given direction. For an OA medium, the optical gyration vector is introduced as a small correction to this equation:
\begin{equation}
    \left(n^2-n_1^2\right)\left(n^2-n_2^2\right)=G
\end{equation}
Near an optical axis, where $n_1\approx n_2$, this equation can be rewritten in the first order as
\begin{equation}
    n_{\pm}=\bar{n}\pm\frac{G}{2\bar{n}}
\end{equation}
where $\bar{n}$ is the average refractive index, whereas $n_{+}$ and $n_{-}$ are the refractive indices for the two circular polarizations. The same result can be obtained using various other formalisms, for example using the Berreman matrices \cite{Berreman1972}. The difference of the refraction indices leads to the rotation of the polarization plane of linearly-polarized light, which is the most well-known signature of OA. This rotation is usually characterized by the optical activity coefficient $\alpha$ expressed in degrees/mm or in rad/mm. Its link with difference in the refractive indices $n_+$ and $n_-$ is given by by the formula $\Delta n=\alpha\lambda/\pi$ where  $\lambda=2\pi/k_0$. Here, $k_0$  is the total wavevector of light composed of its in-plane ($k$ , used in the main text) and vertical ($k_z$) projections $k_0=\sqrt{k_z^2+k^2}$. The OA coefficient can be written in terms of gyration vector $G$ as $\alpha=\pi G/\lambda \bar{n}$ and in terms of gyration tensor $\eta_{ij}$ (usually called $g$, but here we use $\eta$ not to be confused with the quantum geometric tensor of the main text) as $\alpha=\pi \eta_{ij}N_i N_j/\lambda \bar{n}$ ($N_i$ are the direction cosines). These different representations of OA are used in different fields: the OA coefficient $\alpha$ is used to characterize the angle of rotation in the transmission configuration, whereas the gyration tensor $\eta$, being a part of the dielectric permittivity tensor $\epsilon$, is used for the calculation of confined optical modes, like in our case. 

We first discuss the possibility of the OA in perylene crystals. The unit cell is shown in Supplementary Figure~3. The space group of perylene $2_1/m$ contains a screw axis $2_1$ (shown in green) and a perpendicular glide plane (shown in magenta). The molecules of the unit cell of course respect both symmetry elements. However, a delocalized exciton \cite{Rangel2018} with a hole (red) and an electron (blue) occupying different molecules creates a dielectric polarization vector $\vec{P}$ (red). This vector, together with the inhomogeneous charge distribution, break the glide plane symmetry: an electron occupies the place which would correspond to the shifted mirror image of a hole if the glide plane were not broken. Removing the glide plane leads to the allowance of the OA. The fact that the OA is linked with the exciton resonance is confirmed by the inversion of the sign of the effect in Fig.~1(B) of the main text. We note that while optical activity has been traditionally associated with chiral structures and enantiomeres, like tartaric acid, it is actually a much more widespread effect. For example, it has been observed for excitons in ZnSe quantum wells \cite{Kotova2016} and for non-chiral metamaterials \cite{Plum2009}.

Next, we show how the effective Zeeman splitting linear versus in-plane wave vector arises from the OA. As can be seen from the Supplementary Figure~2, the crystallographic axes of perylene are not aligned with the axes of the cavity. The axes $k_x$ (in-plane) and $k_z$ (the quantization direction) are therefore not the main axes of the gyration tensor $\eta_{ij}$. In our case, $k_z=p\pi/L_c$, where $L_c$ is the cavity thickness and $p$ the order of the confined mode ($p=6$ for the mode we consider in the main text). $k_x$  is small in front of  $k_z$, and we can expect the leading terms to appear from the off-diagonal elements of the type $\eta_{xz}k_x k_z/k_0^2$, because all other terms have a higher order in $k_x$. If $k_z\approx k_0$, this expression reduces to $\eta_{xz}k_x/k_0$.

The energy of a beam propagating in an OA media in one of the two circular polarizations can be obtained as $E_{\pm}=\hbar c k_0/n_{\pm}$, where $n_{\pm}=\bar{n}\pm \eta_{xz}k_x /2\bar{n}k_0$ which gives to the first order
\begin{equation}
    E_{\pm}=\frac{\hbar c k_z}{\bar{n}}\mp \frac{\hbar c \eta_{xz}k_x}{2\bar{n}^3}
\end{equation}
Thus, the coefficient of the effective Hamiltonian from the main text is $\zeta=\hbar c \eta_{xz}/2\bar{n}^3$. The gyration tensor value corresponding to the experimentally measured dispersion is $\eta_{xz}\approx 0.14$, much larger than any of the tensor components of quartz $\eta_Q\sim 10^{-5}$, which guarantees that our observations are not due to the optical activity of quartz, a layer of which is present inside the microcavity. At the anticrossing wavelength $\lambda\approx 500$~nm, the splitting between the energies is $E_+-E_-\approx 50$~meV, which gives $\alpha=500$~rad/mm or $\alpha=2.8\times 10^4$~degrees/mm. This is comparable with the values observed in metamaterials, such as chiral photonic crystals \cite{Takahashi2013}.
It starts to approach the optical activity of chiral stacks of 2D materials, where the rotation of tens of degrees can be observed for only 10 monolayers of material, and the corresponding $\alpha_{chir}\sim 5\times 10^6$ degrees/mm \cite{Poshakinskiy2018}.

The optical activity is an effect which stems from non-locality \cite{Landau8}. As such, it can not only originate from the internal properties of a homogeneous material (such as the chirality of an excitonic resonance), but also from inhomogeneities in non-chiral structures, for example, from a transverse thickness gradient \cite{Shelykh2018}. We have made all possible efforts to exclude the possibility that the optical activity observed in our experiments stems from this origin. For this, on the one hand, we have measured the thickness gradient of the structure experimentally, obtaining an average value of $\approx 1.3$~nm/$\mu$m (see the Supplementary Figure 4). On the other hand, we have estimated the energy splitting which could stem from such gradient analytically and numerically, using full 3D electrodynamic simulations in COMSOL(R), and obtained no measureable splitting. To play any role in the effect, the gradient has to be at least 10 times higher. We conclude that we have excluded all alternative explanations of the effect, which therefore must be due to the optical activity of the perylene itself.

\bibliographystyle{naturemag}
\bibliography{scibib}

\begin{addendum}
\item This work was supported by the National Natural Science Foundation of China (Grant No. 91333111, 21503139, 21673144, 21873065, 21833005 and 11804267), the Beijing Natural Science Foundation of China (Grant No. 2162011, 2192011), Natural Science Basic Research Plan in Shaanxi Province of China (Grant No. 2018JQ6041), the High-level Teachers in Beijing Municipal Universities in the Period of 13th Five–year Plan (Grant No. IDHT20180517 and CIT\&TCD20180331), the Open Fund of the State Key Laboratory of Integrated Optoelectronics (IOSKL2019KF01), Capacity Building for Sci-Tech Innovation-Fundamental Scientific Research Funds (025185305000/210, 009/19530050162), Youth Innovative Research Team of Capital Normal University (009/19530050148), Beijing Advanced Innovation Center for Imaging Technology, the Ministry of Science and Technology of China (Grant No. 2013CB933500 and 2017YFA0204503).
We acknowledge the support of the project "Quantum Fluids of Light"  (ANR-16-CE30-0021), of the ANR Labex Ganex (ANR-11-LABX-0014), and of the ANR program "Investissements d'Avenir" through the IDEX-ISITE initiative 16-IDEX-0001 (CAP 20-25).
\item[Author contributions] J.~Ren -- investigation, formal analysis, visualization, methodology, writing; Q. Liao -- conceptualization, funding acquisition, methodology, resources, supervision; F. ~Li -- conceptualization, funding acquisition, methodology, supervision, writing, project administration; Y. ~Li --methodology, writing; H. ~Fu -- conceptualization, funding acquisition, methodology, resources, supervision; J. ~Yao -- funding acquisition, supervision. D.~Solnyshkov -- conceptualization, funding acquisition, formal analysis, methodology, visualization, writing; G.~Malpuech -- conceptualization, funding acquisition, methodology, writing, supervision; O.~Bleu -- conceptualization, validation, methodology, visualization, writing.
 \item[Competing Interests] The authors declare that they have no
competing financial interests.
 \item[Correspondence] Correspondence
should be addressed to liaoqing@cnu.edu.cn (Q.L); felix831204@xjtu.edu.cn (F.L.); hbfu@cnu.edu.cn (H.F.); dmitry.solnyshkov@uca.fr (D.S.) .
\end{addendum}

The datasets generated and analyzed during the current study are available via the Open Science Framework (OSF) repository. The materials are available upon reasonable request.

\clearpage

\section{Supplementary Materials}
\subsection{Supplementary Figures}
\renewcommand{\thefigure}{S\arabic{figure}}
\setcounter{figure}{0}

\clearpage

\begin{figure}[tbp]
\includegraphics[width=1\linewidth]{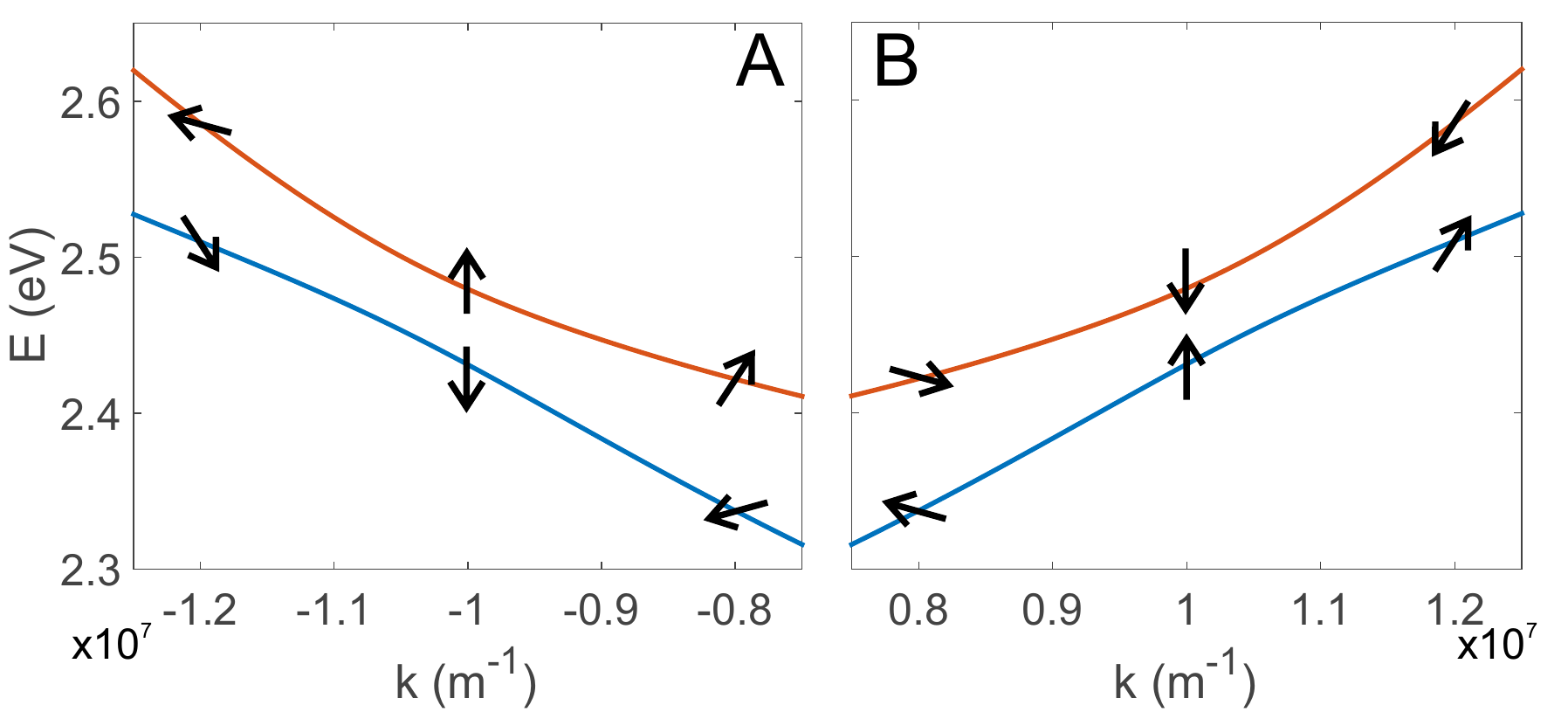}
\caption{\textbf{Eigenstates at the anticrossing.} Dispersion of the upper and lower bands ($E_0$ and $E_1$ from Fig.~2E) around the two anticrossing points (A and B). The pseudospin of the eigenstates is shown with black arrows. Opposite circular polarization is observed for opposite wave vectors, while the in-plane component demonstrates convergent and divergent textures. \label{figS4}
}
\end{figure}

\begin{figure}[tbp]
\includegraphics[width=1\linewidth]{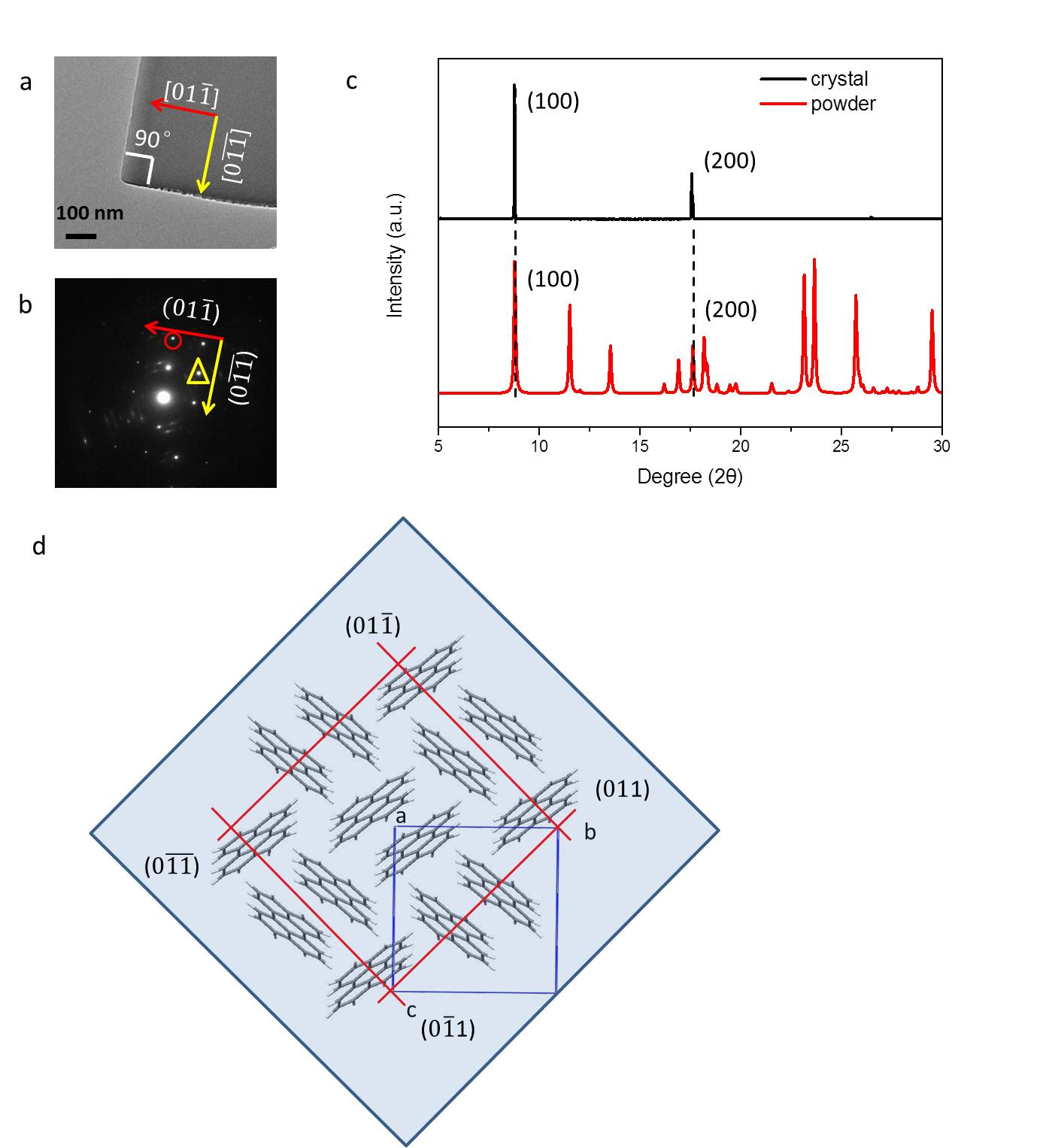}
\caption{\textbf{Material structure.} A and B: TEM image (A) and the corresponding SAED pattern (B) of the perylene microcrystal; C: XRD pattern of the perylene microcrystal (upper panel, black line) and the perylene powder used to fabricate the microcrystal (lower panel, red line); D: The simulated molecule arrangement of perylene from top view. The straight colored lines label the crystal faces (red line) and the unit cell (blue line).\label{figS1} }
\end{figure}

\begin{figure}[tbp]
\centering
\includegraphics[width=0.7\linewidth]{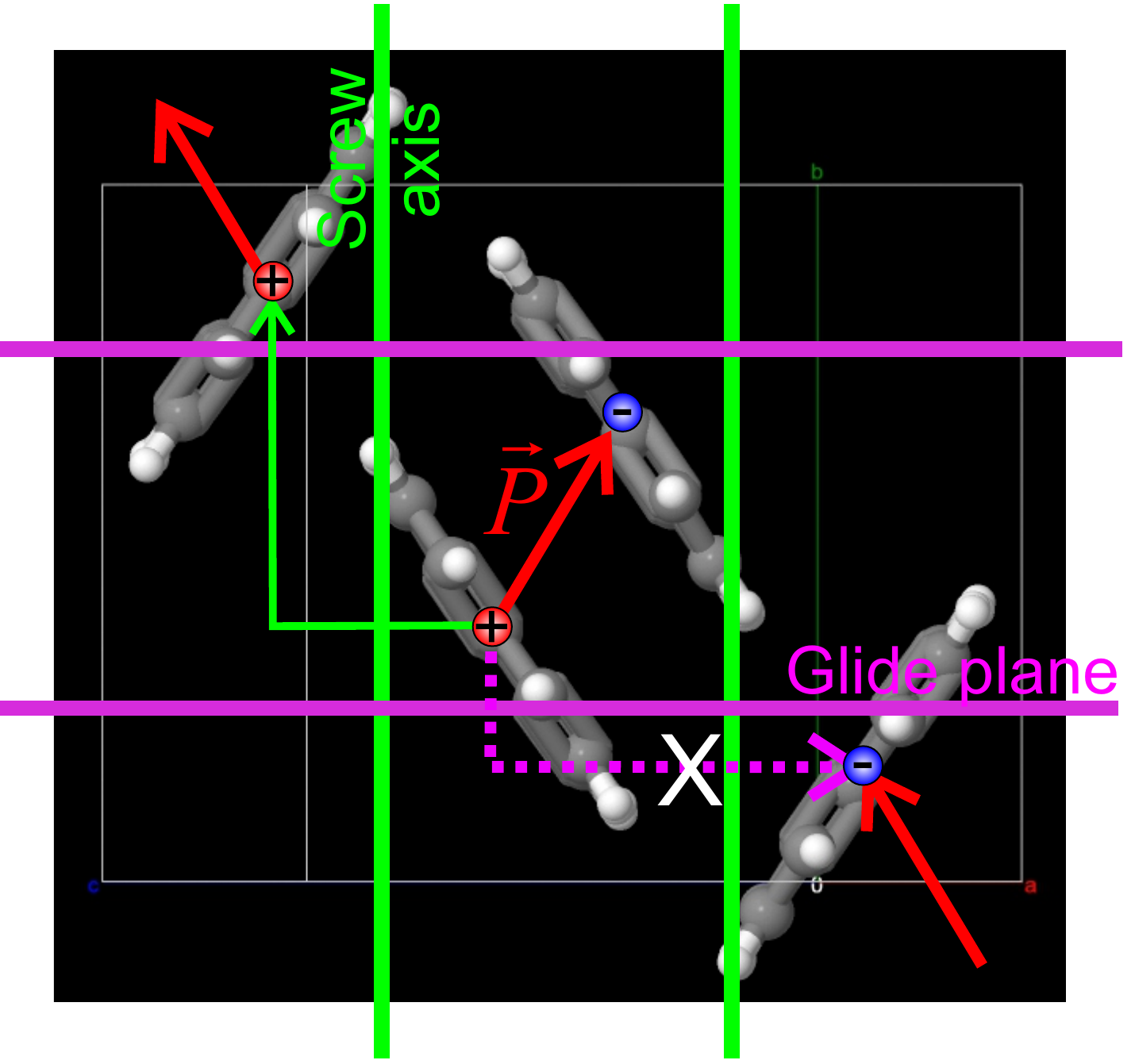}
\caption{\textbf{Excitons in perylene.} The unit cell of perylene with its symmetry elements: screw axis (green) and glide plane (magenta) with the corresponding operations (green and magenta arrows). The charges composing the delocalized exciton are shown in their respective locations, with the dielectric polarization vector $\vec{P}$ (red). The distribution of charges and the dielectric polarization violate the glide plane symmetry (white cross), making the OA possible. \label{figS0}
}
\end{figure}

\begin{figure}[tbp]
\centering
\includegraphics[width=0.7\linewidth]{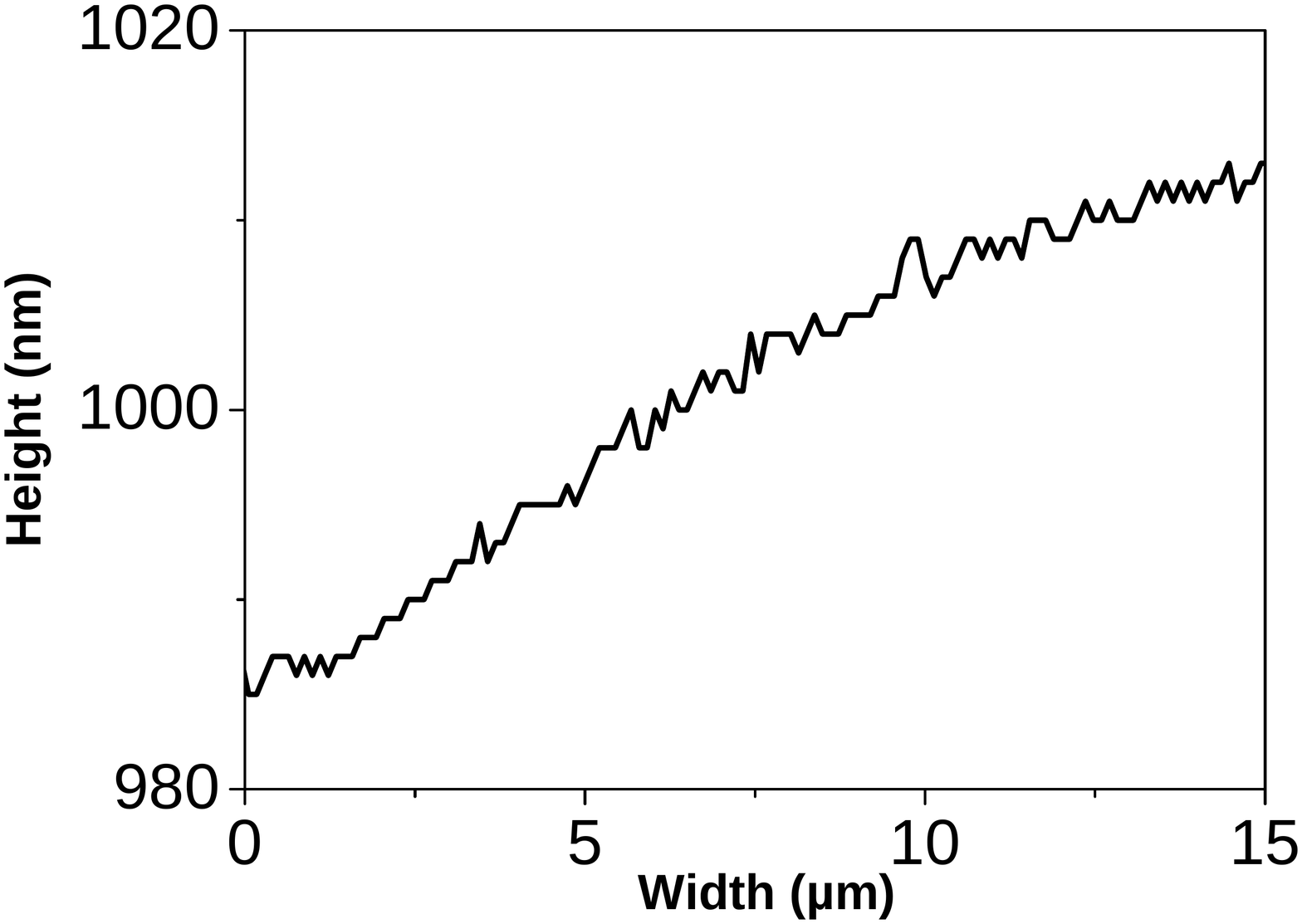}
\caption{\textbf{Cavity thickness variation.} Experimentally measured cavity thickness exhibiting a gradient of $\approx 1.3$~nm/$\mu$m for the whole structure.}
\end{figure}

\end{document}